\documentclass[prc,twocolumn,superscriptaddress,showpacs,twoside,floatfix]%
{revtex4-1}

\usepackage{amssymb,epsfig}

\begin{document}

\title{Anomalous population of $^{10}$He states in reactions with $^{11}$Li}

\author{P.G.~Sharov}
\affiliation{Flerov Laboratory of Nuclear Reactions, JINR, Dubna, RU-141980
Russia}
\affiliation{SSC RF ITEP of NRC ``Kurchatov Institute'', Moscow 
RU-117218 Russia}
\author{I.A.~Egorova}
\affiliation{Bogoliubov Laboratory of Theoretical Physics, JINR, Dubna,
RU-141980
Russia}
\affiliation{SSC RF ITEP of NRC ``Kurchatov Institute'', Moscow 
RU-117218 Russia}
\author{L.V.~Grigorenko}
\affiliation{Flerov Laboratory of Nuclear Reactions, JINR, Dubna, RU-141980
Russia}
\affiliation{GSI Helmholtzzentrum f\"{u}r Schwerionenforschung,
  Planckstra{\ss}e 1, D-64291 Darmstadt, Germany}
\affiliation{National Research Center ``Kurchatov Institute'', Kurchatov sq.\
1, RU-123182 Moscow, Russia}

\date{\today.}

\begin{abstract}
Structure with the lowest energy observed in the $^{10}$He spectrum populated
in the proton
knockout reaction with $^{11}$Li beam has a peak at $1.2-1.5$ MeV. This peak is
usually interpreted as a resonant $0^+$ ground state of $^{10}$He. Our
theoretical calculations indicate that this peak is likely to be a pileup of
$1^-$, $0^+$, and $2^+$ excitations with very similar shapes.
%We predict a very specific nature of the $1^-$ excitation in $^{10}$He.
Moreover, the ``soft''
$1^-$ excitation appears to be the lowest one in energy. Such an anomalous
continuum response is traced to the halo structure of $^{11}$Li providing
extreme low energy shift to all the expected continuum excitations.
Competitions of the initial state structure (ISS) and the final state
interaction (FSI) effects on the spectrum and three-body correlations in 
$^{10}$He  are discussed. Analogous effect of the extreme low-energy shift
could also
be expected in other cases of $2n$ emitters populated in reactions with halo
nuclei. Simplified example of the $^{10}$He spectrum in $\alpha$ knockout from
$^{14}$Be, is given. We also discuss limits on the properties of $^{9}$He
stemming from the observed $^{10}$He spectrum.
\end{abstract}


\maketitle

%===============================================================================

\section{Introduction}
%
%===============================================================================

The dripline systems often represent complicated and unusual forms of
nuclear structure. These include nucleon halos, soft continuum excitation modes,
abnormal cross sections for certain reactions. The $^{10}$He nucleus has one of
the largest known $N/Z$ ratio representing the extreme situation of the nuclear
matter asymmetry. This system is very complicated for studies and due to neutron
excess there are only few realistic methods to populate states in this
interesting nuclear system.

Since the experimental discovery of $^{10}$He in 1994 \cite{Korsheninnikov:1994}
there were only several experimental studies
\cite{Ostrowski:1994,Kobayashi:1997,Golovkov:2009,Johansson:2010,Sidorchuk:2012,
Kohley:2012}. Among the available experimental data on $^{10}$He we would like
to skip the discussion of two results \cite{Ostrowski:1994,Kobayashi:1997} 
which are complicated for interpretation. Then we end with two ``lines'' of 
research which we can consider as well established results.
(i) Low-lying peak in the
$^{10}$He spectrum was observed in knockout reactions from halo nuclei: from
$^{11}$Li in Ref.\ \cite{Korsheninnikov:1994} providing $E_T=1.2(3)$ MeV, from
$^{11}$Li in Ref.\ \cite{Johansson:2010} providing $E_T=1.4-1.5$ MeV,
$\Gamma \sim 1.9$ MeV, and from $^{14}$Be ($\alpha$ or $2p$ removal
\cite{Kohley:2012})
giving $E_T=1.60(25)$ MeV, $\Gamma = 1.8(4)$ MeV. It should
be noted that the low-lying spectra in all these experiments are practically
identical, although the details of data interpretation are different. We
demonstrate in this work that it could be more than just a coincidence.
(ii) $^{10}$He spectrum was populated by the $^3$H($^{8}$He,$p$) two-neutron
transfer reaction in two consequent works \cite{Golovkov:2009,Sidorchuk:2012}.
In Ref.\ \cite{Sidorchuk:2012} the cross section peak with $E_T=2.1(2)$ MeV and
$\Gamma \sim 2$ MeV was observed. Also an evidence was obtained in this work
for anomalous $\{0^+,1^-,2^+ \}$ level ordering in $^{10}$He. The earlier
experiment \cite{Golovkov:2009} has a very limited statistics to give a
quantitative result, but the conclusion, that there is no indication of a
resonant $^{10}$He state at about $1.0-1.5$ MeV, is valid in both works. Thus 
for the first glance there is a contradiction between two types of the data.

Theoretical results on $^{10}$He are also sparse. The $^{10}$He ground state
structure and decay were studied in papers
\cite{Aoyama:2002,Volya:2007,Grigorenko:2008} providing results in generally
consistent, but with a lot of differences in details. In
Ref.~\cite{Grigorenko:2008} broad exploratory studies of the $^{10}$He g.s.\
population were performed in the framework of the cluster $^{8}$He+$n$+$n$
model. The situation with interactions in the $^{8}$He+$n$ channel was quite
uncertain at that moment due to controversial experimental data. This led to
uncertain calculation results and essential freedom in interpretations. From
theoretical point of view the calculations of $^{10}$He spectrum is a reflection
of the $^{9}$He low-lying spectrum properties. Since that time we have got a
confidence that in reality the spectrum of $^{10}$He could be used to impose
limitations on the spectrum of $^{9}$He. This can be made, however, only if the
understanding of the reaction mechanisms leading to population of $^{10}$He is
achieved.

Already in the paper \cite{Grigorenko:2008} it was mentioned about the possible
importance initial state effects for population of the $^{10}$He in reactions
with $^{11}$Li. In paper \cite{Sidorchuk:2012} we extended qualitative
calculations of Ref.\ \cite{Grigorenko:2008} to the $^{10}$He excitations with
different $J^{\pi}$ populated in transfer reactions providing explanation for
experimental data \cite{Sidorchuk:2012}. Extension of our calculations for
population of $^{10}$He excitations with different $J^{\pi}$ in the knockout
reactions required considerably more effort and provided quite unusual results
which we report here. These results well explain the above mentioned
contradiction in the two types of data. They demonstrate potential danger of
simple-minded attitude to interpretation of reaction data with exotic nuclei and
promote the importance of a careful theoretical assessment.

%===============================================================================

\section{Sudden removal of proton and three-body model}

%===============================================================================

In this work we report a large qualitative effect and we presume that even
relatively simple reaction model --- sudden removal of a proton --- is
sufficient at this stage. The model for sudden removal of a proton from
$^{11}$Li populating three-body $^{10}$He continuum can be reformulated as a
solution of the inhomogeneous three-body Schr\"odinger equation
\begin{equation}
(\hat{H}_3 - E_T) \Psi^{JM(+)}_{E_T}(X,Y) = \Phi^{JM}_q (X,Y).
\label{eq:schred}
\end{equation}

In this work we rely on the three-cluster $^{8}$He+$n$+$n$ model of $^{10}$He
and on the set of $^{8}$He-$n$ potentials developed in Ref.\
\cite{Grigorenko:2008}. These potentials give a  purely repulsive $s$-wave with
$a_s \approx 3$ fm and the $p_{1/2}$ ground state of $^{9}$He at 2 MeV and the
$d_{5/2}$ state at about 5 MeV (the properties of the $^{9}$He resonant states
are those found in Ref.\ \cite{Golovkov:2007}). The justification for this
choice is provided in Section~\ref{sec:limits}.

\begin{figure}
\begin{center}
\includegraphics[width=0.45\textwidth]{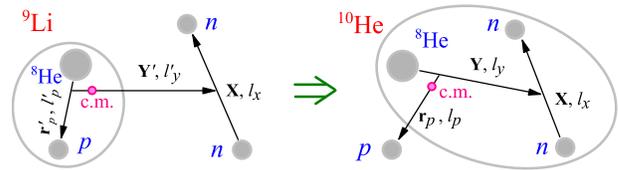}
\end{center}
\caption{The coordinate schemes used for proton removal (cluster removal in the
general case) calculations.}
\label{fig:coord}
\end{figure}

In our theoretical approach the cross section for population of the three-body
continuum is proportional to flux defined via WF with outgoing wave asymptotic:
\[
\frac{d \sigma}{d E_T} \sim j(E_T) = \frac{1}{M} \mathop{Im} \! \int d \Omega_5
\Psi^{(+)\dagger}_{E_T}\rho^{5/2} \! \! \left. \frac{d}{d\rho} \rho^{5/2}
\Psi^{(+)}_{E_T} \right|_{\rho_{\max}}.
\]
The three-body $^{10}$He WF $\Psi^{(+)}_{E_T}$ found in the hyperspherical
harmonic method, depends either of hyperspherical variables $\{\rho,\Omega_5\}$,
see Ref.\ \cite{Grigorenko:2008} for details, or on the equivalent set of the
Jacobi vectors $\{\mathbf{X},\mathbf{Y} \}$, see Fig.~\ref{fig:coord}.

To define the source function  $\Phi^{JM}_q (X,Y)$ we use the cluster $^{11}$Li
WF in the form
\begin{equation}
\Psi^{J_{i}M_{i}}_{^{11}\text{Li}}(\mathbf{X},\mathbf{Y}',\mathbf{r}_p')=
[\Psi^{(3b)}_{^{11}\text{Li}}(\mathbf{X},\mathbf{Y}') \otimes
\Psi_{^{9}\text{Li}}(\mathbf{r}_p') ]_{J_{i}M_{i}} \,,
\label{eq:psi-init}
\end{equation}
combining the three-body cluster WF of $^{11}$Li $\Psi^{(3b)}_{^{11}\text{Li}}$
with the single particle WF of the proton motion within the $^{9}$Li cluster
$\Psi_{^{9}\text{Li}}$. The three-body WF is taken from paper
\cite{Shulgina:2009} where it is constructed in the analytical Ansatz by fitting
various observables to the available experimental data. We used the
simplest form of the WF from \cite{Shulgina:2009} (without coupling to the spin
on the core).

The $p$-wave $^{8}$He+$p$ WF $\Psi_{^{9}\text{Li}}(\mathbf{r}_p')$ is
constructed in the potential model. We use the Woods-Saxon potential with the
depth $-56$ MeV radius 3.03 fm and diffuseness 0.75 fm.
It provides experimental binding energy for proton $^9$Li ($E_b= 13.933$ MeV)
and root mean square (rms) radius $\langle r_p' \rangle = 2.89$ fm, which
allows to reproduce the rms matter radius of $^{9}$Li with experimental rms
matter radius of the $^{8}$He cluster. The momentum space representation namely
the squared formfactor of $p+^8$He WF is illustrated in Fig.~\ref{fig:formfac}.

%-------------------------------------------------------------------------------
%-------------------------------------------------------------------------------

Illustration of the reaction mechanism and relevant variables is provided by
Fig.~\ref{fig:coord}. In the initial coordinate system the sudden removal of
proton from $^9$Li core lead to momentum transfer to the $^{8}$He cluster of the
$^{9}$Li core. To define the effect of such a momentum transfer on the whole
final WF of $^{10}$He we perform the coordinate transformation
$\{\mathbf{Y}',\mathbf{r}_p'\} \rightarrow \{\mathbf{Y},\mathbf{r}_p \}$ of WF
(\ref{eq:psi-init}). This is accomplished by decomposition of
$\Psi^{J_{i}M_{i}}_{^{11}\text{Li}}$ over hyperspherical harmonics for fixed $X$
and
$l_x$ values of the two neutron subsystem and use of the Raynal-Revai
transformation \cite{Raynal:1970}. This transformation leads to WF in which
proton coordinate is connected with the whole system ($^{10}$He) center-of-mass.
In new coordinates $\{\mathbf{Y},\mathbf{r}_p \}$ the source term for $^{10}$He
production is provided by acting on the proton by the annihilation operator. In
coordinate space this operation gives
\begin{equation}
\Phi_{\mathbf{q}} (\mathbf{X},\mathbf{Y}) = \int d^3 \mathbf{r}_p e^{i
\mathbf{q}\mathbf{r}_p} \Psi_{^{11}\text{Li}}
(\mathbf{X},\mathbf{Y},\mathbf{r}_p).
\label{eq:source-vec}
\end{equation}
The source functions for different $J^{\pi}$ in $^{10}$He are obtained by
angular momentum decomposition
\begin{eqnarray}
\Phi^{JM}_{q,\gamma,l_p} (X,Y) = \int  d \Omega_x d \Omega_y  d \Omega_q
\, \Phi_{\mathbf{q}} (\mathbf{X},\mathbf{Y}) \nonumber \\
\times [[[Y_{l_x}(\hat{X}) \otimes Y_{l_y}(\hat{Y}) ]_L \otimes \chi_S ]_J
\otimes Y_{l_p}(\hat{q}) ]_{JM}.
\label{eq:source-part}
\end{eqnarray}
The multi-index $\gamma=\{LSl_xl_y\}$ defines the complete set of angular quantum
numbers for the core+N+N three-body problem.

For exploratory model calculations we also use the sources defined by the simple
analytical expression:
\begin{eqnarray}
\Phi^{JM}_{q,\gamma} (X,Y)= \sum_{K,\gamma} N_{K,\gamma} \,
\frac{f(\rho)}{\rho^{5/2}} \,
\mathcal{J}_{K \gamma}^{JM}(\Omega_{\rho}), \nonumber \\
f(\rho) = \sqrt{7/5} (21\rho^{5/2}/\rho_0^3) \,
\exp[-\sqrt{21/2}\,(\rho/\rho_0)].
\label{eq:source-exp}
\end{eqnarray}
The function $f(\rho)$ is defined in such a way that it is normalized and its
rms radius is defined by $\langle \rho \rangle= \rho_0$.

%===============================================================================

\section{Properties of the sources}
%
%===============================================================================

To understand the continuum calculations of this work we need to illustrate the
properties of the obtained source functions in details.

%-------------------------------------------------------------------------------
\begin{figure}
\begin{center}
\includegraphics[width=0.42\textwidth]{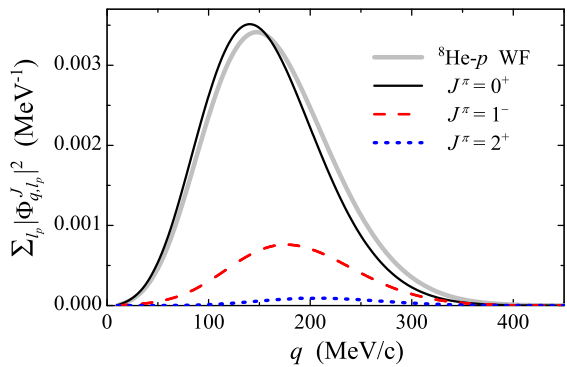}
\end{center}
\caption{Squared formfactors for population of different states in $^{10}$He.
Realistic $^{9}$Li WF\@. The proton formfactor for $^{9}$Li WF is shown for
comparison by thick gray curve. It is scaled to match the normalization of the
$0^+$ formfactor.}
\label{fig:formfac}
\end{figure}
%-------------------------------------------------------------------------------

In our treatment of the reaction with $^{11}$Li we assume the removal of the
$p_{3/2}$ proton from the  $^{9}$Li core leading to the $^{8}$He g.s.\
population. The integrated $^{10}$He formfactors, as function of the momentum
transfer $q$
\begin{equation}
|\Phi^{JM}_{q,\gamma,l_p}|^2 = \int d\Omega_q \, d^3 X \, d^3 Y \, |
\Phi_{\mathbf{q}}
(\mathbf{X},\mathbf{Y}) |^2,
\label{eq:formfac}
\end{equation}
are shown in Fig.~\ref{fig:formfac}. The formfactors are normalized as
\begin{equation}
\sum_{\gamma l_p} \int dq \,|\Phi^{JM}_{q,\gamma,l_p}|^2 = \frac{\pi}{2},
\label{eq:formfac-norm}
\end{equation}
for $^{11}$Li WF normalized to unity.  The Fig.~\ref{fig:formfac} provides the
first idea about population of $^{10}$He states with different $J^{\pi}$ in our
model: it is clear that beside $0^+$ population we can expect a sizable
population of the $^{10}$He continuum with $1^-$ and $2^+$ states.
The momentum distribution of the $0^+$ formfactor is very similar to
initial formfactor of the $^{8}$He-$n$ motion, but for higher $J^{\pi}$ there is
a significant shift to higher momenta.

\begin{figure}
\begin{center}
\includegraphics[width=0.47\textwidth]{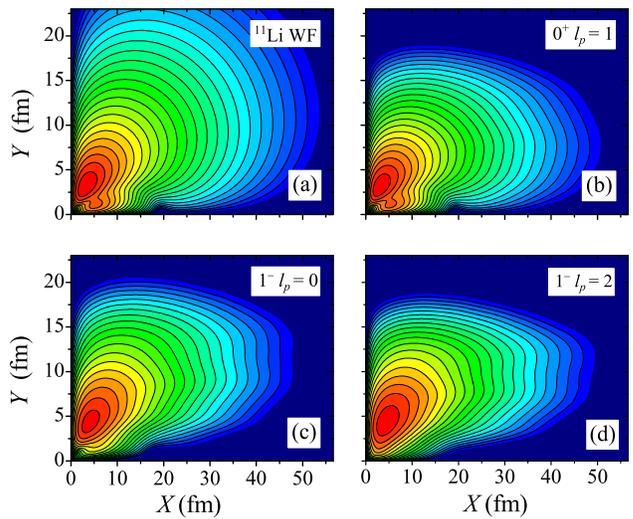}
\end{center}
\caption{Correlation densities for $^{11}$Li g.s.\ WF (a) and for different
source functions for $^{10}$He (b,c,d). The $q$ values chosen for different
$J^{\pi}$ correspond to peaks of the formfactors, see Fig.~\ref{fig:formfac}.
The scale is four contours per order of the magnitude.}
\label{fig:corel-dens}
\end{figure}

The correlation densities for the most important $^{10}$He source functions
are illustrated in Fig.~\ref{fig:corel-dens}. For the $0^+$ state the
correlation density is very similar to the one for the initial $^{11}$Li WF
(Fig.~\ref{fig:corel-dens}a) when we look on the bulk of the WF\@. In asymptotic
regions significant differences can be seen.

%-------------------------------------------------------------------------------
%-------------------------------------------------------------------------------

\begin{figure}
\begin{center}
\includegraphics[width=0.42\textwidth]{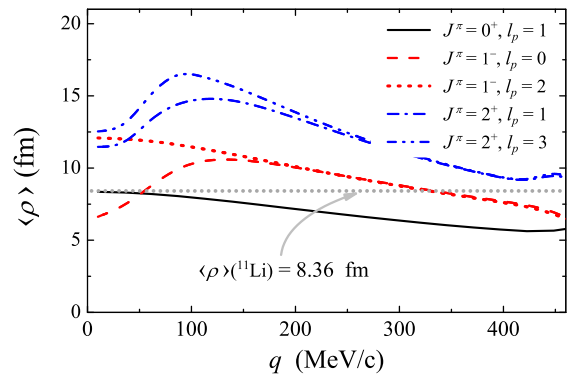}
\end{center}
\caption{Root mean square hyperradii for the source functions of different
states in $^{10}$He. The rms $\rho$ value for the $^{11}$Li WF is shown by gray
dotted line.}
\label{fig:mean-radii}
\end{figure}

The radial properties of source functions are illustrated in
Fig.~\ref{fig:mean-radii}. One aspect of these properties is well expected:
the size of the generated source is decreasing with the increase of the momentum
transfer. 
In the limit of small momentum transfer the rms radius of the $0^+$ source tend
to that of initial $^{11}$Li WF\@. This mean that in the first approximation the
population of the $^{10}$He g.s.\ can be considered as if it is happening
directly
from radial configurations of the $^{11}$Li WF\@. It should be noted that with
the increase of $J$ the radial extent of the sources significantly grows,
which lead
to important consequences for the obtained excitation spectra of $^{10}$He, see
Fig.~\ref{fig:no-fsi-all} and discussion in the Section~\ref{fig:no-fsi-all}.

%-------------------------------------------------------------------------------
%-------------------------------------------------------------------------------

%===============================================================================
\begin{table}[b]
\caption{Source function properties for calculations of
  Fig.~\ref{fig:no-fsi-all} obtained by variation of the radial size of WF in
  the
$^{8}$He-$p$ channel. This variation was achieved by varying potential depth and
thus the proton binding energy $E_b$. Probabilities for different $J^{\pi}$
population are in percent.}
\begin{ruledtabular}
\begin{tabular}[c]{cccccc}
case  & $\langle r_p' \rangle$ (fm) & $E_b$ (MeV) & $W(0^+)$ & $W(1^-)$ &
$W(2^+)$  \\
\hline
``Narrow'' & 2.0  & -30.   & 66.3 & 27.6 & 6.2 \\
Realistic  & 2.86 & -13.93 & 79.2 & 18.5 & 2.3 \\
``Broad''  & 3.5  & -0.5   & 84.6 & 14.2 & 1.1 \\
\end{tabular}
\end{ruledtabular}
\label{tab:var-9li}
\end{table}
%===============================================================================

% term. CM recoil should be replaced
% In the proposed reaction model the population of configurations with different
% $J^{\pi}$ beside $0^+$ is due to the center-of-mass recoil effect for
%$^{10}$He
% compared to initial $^{11}$Li center-of-mass.
In the proposed reaction model the $^{10}$He recoil, the energy transfer into
the $^{10}$He system, and the $\Delta J^\pi$ transfer to $^{10}$He are taken to
account. $\Delta J^\pi$ should be mainly sensitive to two aspects of the
structure:
(i) the mass difference between the removed particle (proton) and the core of
the final system ($^{8}$He),
(ii) the relative size of the orbital motion of the removed particle and halo
particles.

Point (i) can be used for consistency checks: in the limit $M(^8$He$)\to
\infty$ only the $J^{\pi}=0^+$ source survives, while the others tend to zero.
The actual scale of this trend is a such that for $M(^8$He$)\gtrsim 30 M_N$ the
configuration with $J^{\pi}=0^+$ is populated with $99\%$ probability.

Point (ii) is illustrated in Fig.~\ref{fig:no-fsi-all}. These figures show
excitation spectra of $^{10}$He obtained by the Fourier transformation of
sources obtained with $^{8}$He-$p$ relative motion WFs obtaining by variation
of radial extent. Fig.~\ref{fig:no-fsi-all}(b) corresponds to realistic WF,
while the (a) and (c) cases should demonstrate a possible variation scale. Such
excitation spectra correspond to the physical situation of no FSI in the
outgoing $^{10}$He channel. This can be considered as an extreme test case of
initial state structure effects on the observable $^{10}$He properties.
Following main features of the distributions in Fig.~\ref{fig:no-fsi-all}
should be emphasized:

%-------------------------------------------------------------------------------
\begin{figure*}
\begin{center}
\includegraphics[width=0.326\textwidth]{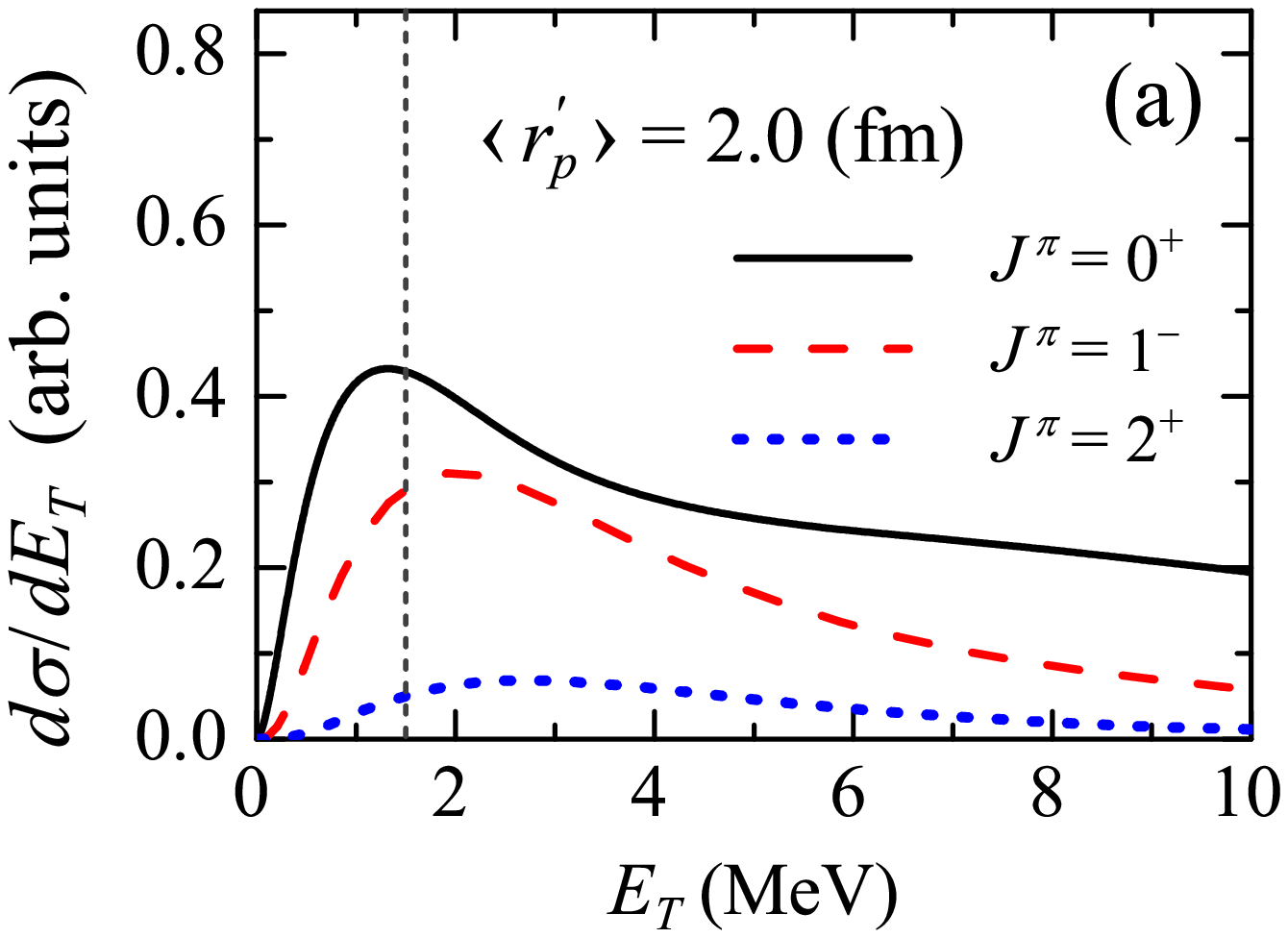}
\includegraphics[width=0.27\textwidth]{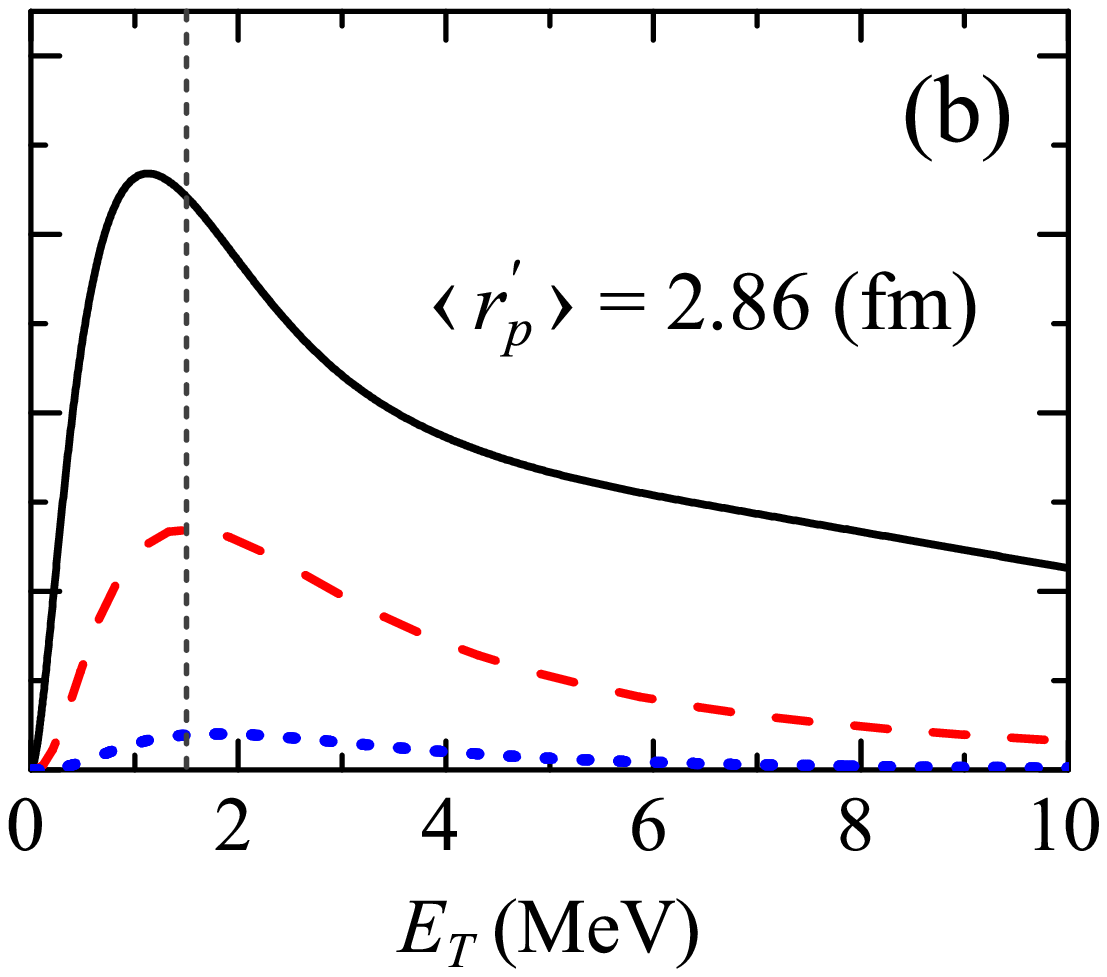}
\includegraphics[width=0.27\textwidth]{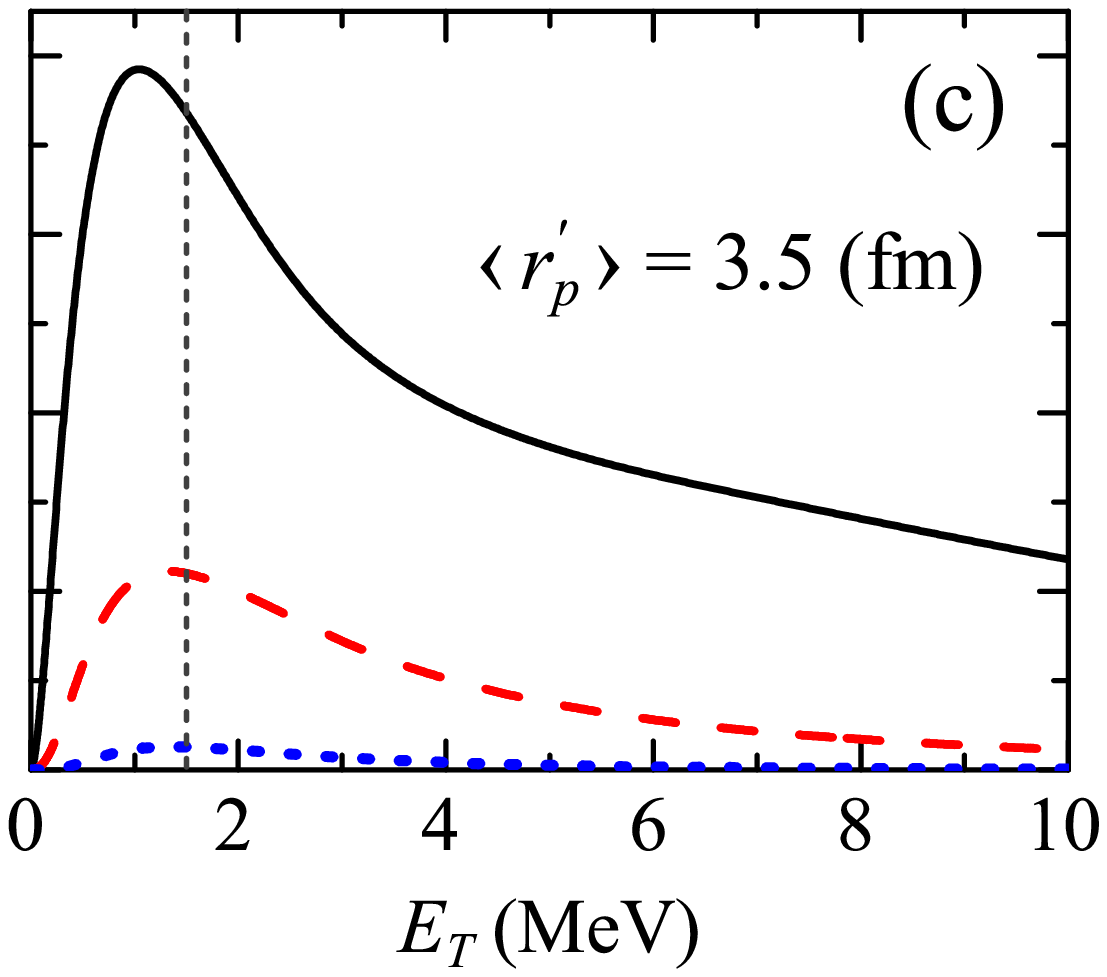}
\end{center}
\caption{Fourier transform for $^{10}$He sources provides the excitation
functions for $0^+$, $1^-$, and $2^+$ in the approximation of no final state
interactions. Different panels show the results obtained for different
$^{8}$He+$p$ WFs. The calculations with realistic rms radius $\langle
r_p'\rangle$
are given in panel (b), while the results of (a) and (c) demonstrates the
possible scale of variation. Vertical dotted line at 1.5 MeV is shown to guide
the eye.}
\label{fig:no-fsi-all}
\end{figure*}
%-------------------------------------------------------------------------------

\begin{enumerate}

\item Even very large variation of the $^{11}$Li radius leads to moderate
variation in the ratio of different $J^\pi$ population.

\item Population of higher $J^{\pi}$ grows with asymmetry between the WF size of
removed particle and in the WF size of the halo configuration of the $^{11}$Li.

\item Population of higher $J^{\pi}$ is never small. It can be expected up to
$50 \%$ for the low-energy excitations of $^{10}$He. However, there also could
hardly be expected less than $20 \%$ of $1^-$ population.

\item Even \emph{without any final state interaction} in the $^{10}$He channel
we can expect very low-energy response in the $^{10}$He spectrum. It is actually
peaked at \emph{lower energy} than observed experimental peak in the spectrum of
$^{10}$He, see Fig.~\ref{fig:spec-j-pi}(c).

\item Excitations with higher $J^{\pi}$ without FSI have profiles very similar
to those of the $0^+$ contribution with peaks mainly below $E_t\sim 2$ MeV of 
excitation in $^{10}$He.

\end{enumerate}

Having in mind this anomalous behavior of the sources, let's turn to $^{10}$He
dynamical three-body continuum calculations.

%===============================================================================

\section{Spectra of $^{10}$He states}

%===============================================================================

\begin{figure}
\begin{center}
\includegraphics[width=0.38\textwidth]{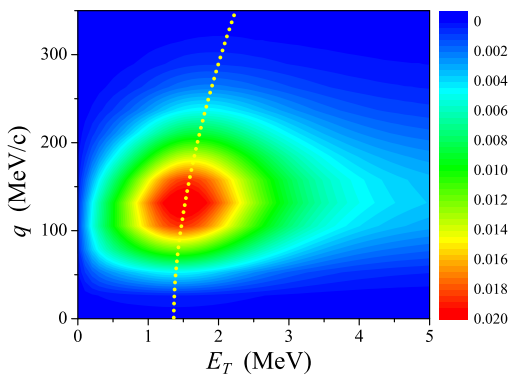}
\end{center}
\caption{Contour plot of the population cross section for $^{10}$He $0^+$ g.s.\
as function of the decay energy $E_T$ and the transferred momentum $q$. The
dotted curve is provided to guide the eye for position of the cross section
peak as function of $q$.}
\label{fig:q-vs-e}
\end{figure}

The spectrum of $^{10}$He ($0^+$) is calculated with the source function Eq.\
(\ref{eq:source-part}), is shown in Fig.~\ref{fig:q-vs-e} as a function of the
transferred momentum $q$. At very low transferred momentum, e.g.\ $q<50$ MeV/c,
the sanctum is very stable. With the increase $q$ the spectrum shifts to
considerably higher energy $E_T$: by $\sim 250$ keV for peak values and by $\sim
1$ MeV for $q$ approximately 300 MeV/c. This effect should be taken into
account thinking about more realistic reaction calculations. It is known that
in the Glauber model the reaction is more peripheral and the momentum transfer
is shifted towards the lower values. Thus some reduction in the $E_T$ peak
position can be expected compared to this calculation.

%-------------------------------------------------------------------------------
%-------------------------------------------------------------------------------

\begin{figure}
\begin{center}
\includegraphics[width=0.40\textwidth]{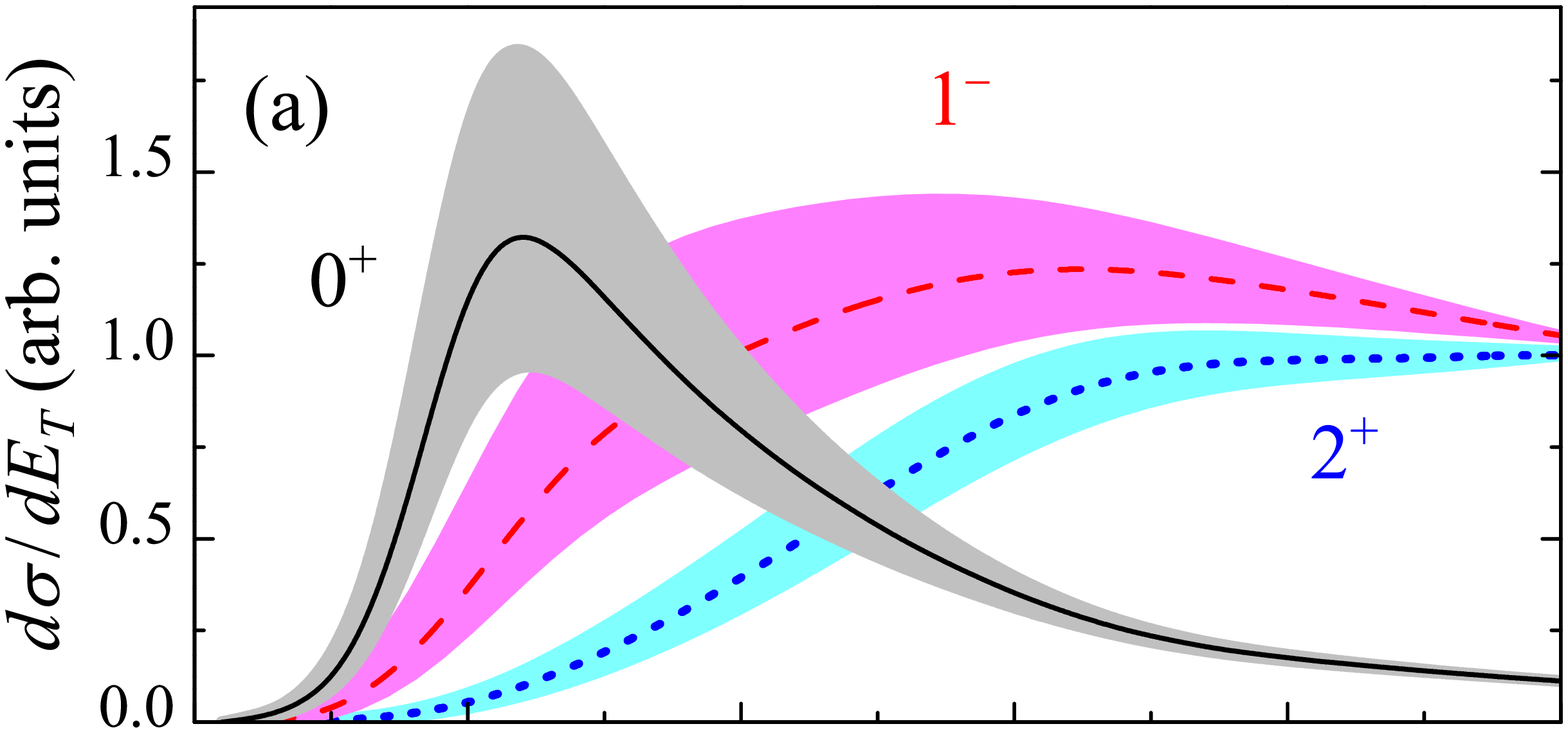}  \\
\includegraphics[width=0.40\textwidth]{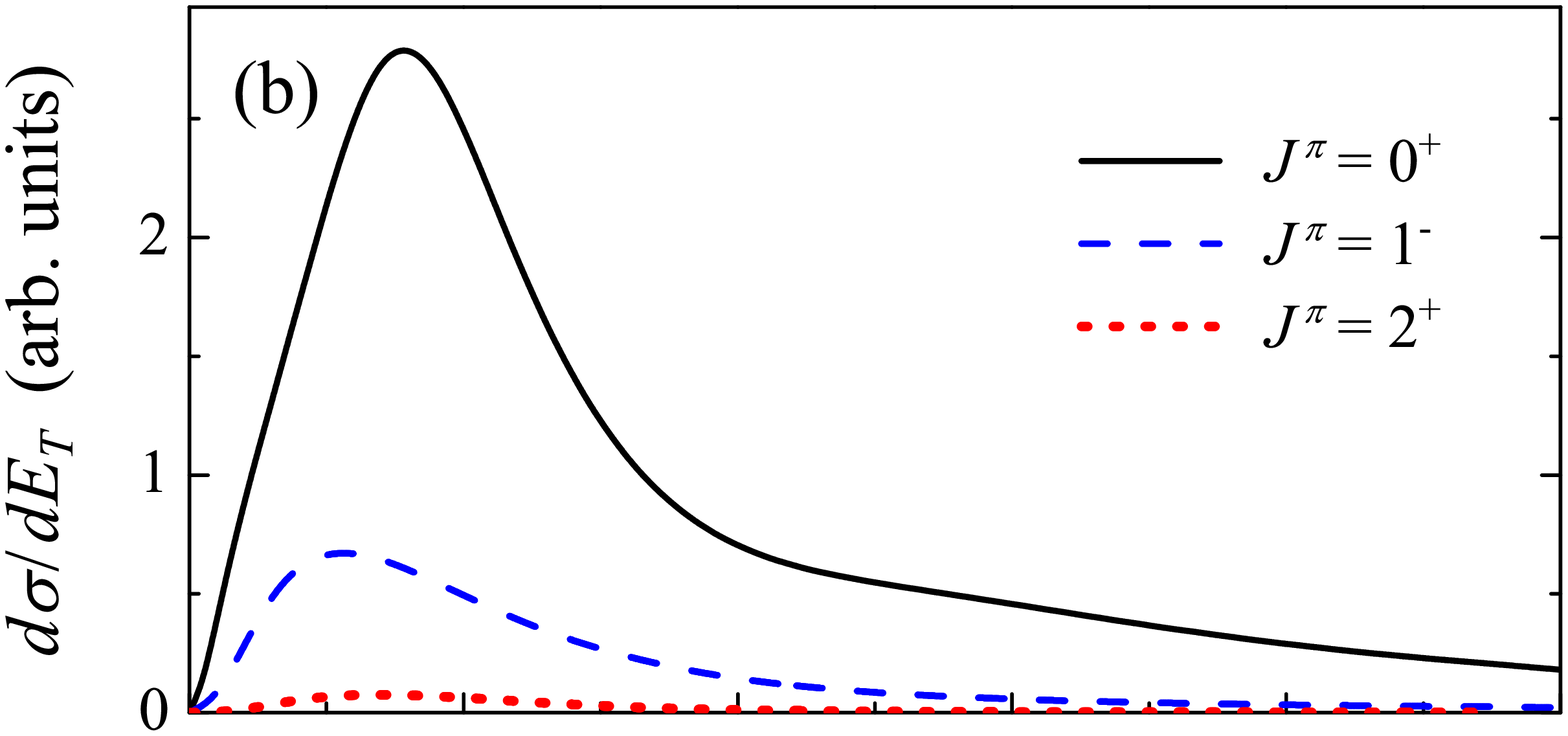}  \\
\includegraphics[width=0.40\textwidth]{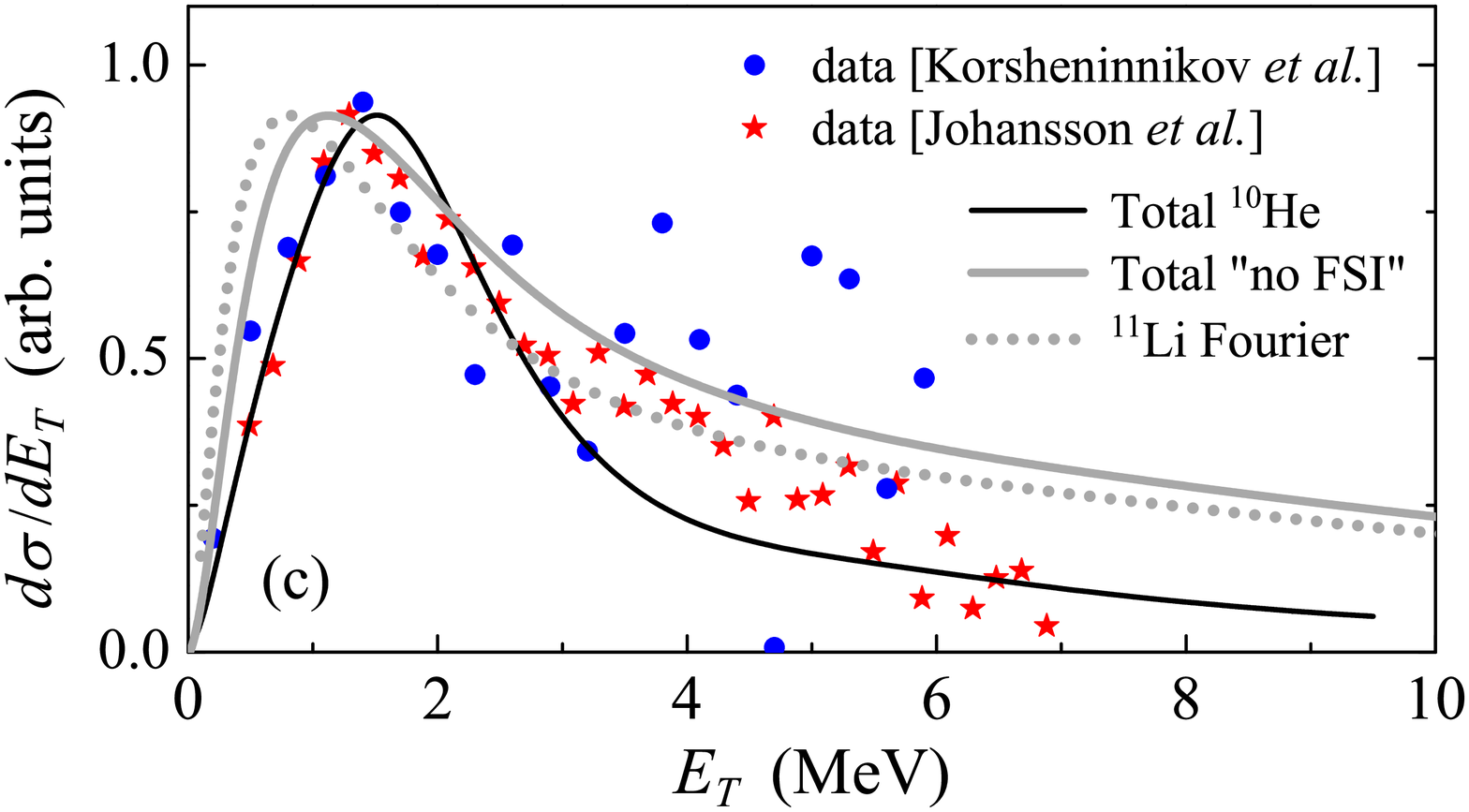}
\end{center}
\caption{The predicted spectra of states with different $J^{\pi}$. Panel (a)
shows estimate results for ``conventional'' situation. Calculations for proton
removal from $^{11}$Li are given in panel (b). The total spectrum is
qualitatively compared with experimental data
\cite{Korsheninnikov:1994,Johansson:2010} in panel (c).
Vertical dotted line at 2 MeV is shown to guide the eye.}
\label{fig:spec-j-pi}
\end{figure}

For conventional situation of nuclear reactions we expect the source size to be
of the typical nuclear size [$\langle \rho \rangle = 3-5$ fm]. The calculation
with the source Eq.\ (\ref{eq:source-exp}) with such radii shown in
Fig.~\ref{fig:spec-j-pi} (a).

The situation changes drastically when we turn to the $^{11}$Li source
[see Fig.~\ref{fig:spec-j-pi} (b)]. All three major configurations predicted
to be populated  $\{0^+,1^-,2^+\}$ are shifted now to extreme low energy
$E_T \sim 1.3-1.6$ MeV. The overall picture is in a close analogy with what can
be seen in Fig.~\ref{fig:no-fsi-all} for calculations without any FSI\@. It is
evident that the ``momentum content'' of the sources associated with nucleon
removal from $^{11}$Li is such low that the observed excitation spectrum is not
significantly sensitive to the FSI any more and reflects predominantly the
initial state structure. What especially attracts the attention here is that
the states with different $J^{\pi}$ (the $0^+$, $1^-$, and $2^+$
configurations) are peaked at practically the same energy, which strongly
contradicts typical expectations that they are separated by some MeVs.

Fig.~\ref{fig:spec-j-pi} (c) provides a qualitative comparison of spectra
provided by different models with two available data sets for proton removal
from
$^{11}$Li \cite{Korsheninnikov:1994,Johansson:2010}. The Fourier transformation
of the $^{11}$Li WF provides the excitation spectrum for $^{10}$He peaked at too
low energy which immediately can be ruled out. The accounting for the reaction
mechanism in our model but without FSI leads to the results much closer to
experiment. The inclusion of FSIs bring us to a better agreement with the
experiment. We should pay an attention here that the calculated effect of FSI is
opposite to common expectation: the cumulative effect of the $^{10}$He FSI is
the shift of the spectrum to a higher energy and thus can been seen as an
\emph{effective repulsion}.

%-------------------------------------------------------------------------------
%-------------------------------------------------------------------------------

The reaction mechanism model considered in this work (sudden proton removal)
can be considered as quite simplistic. However, we know that such an approach
provides qualitatively correct results for majority of high-energy fragmentation
reactions with exotic nuclei. Moreover, if a more realistic model (e.g.,
Glauber) is considered then the reaction becomes more peripheral. Our estimates
indicate that in such a case the relative population of the $1^-$ state is
enhanced. Thus our results could be considered as a conservative estimate of the
proposed phenomenon of the anomalous $^{10}$He spectrum population.

%===============================================================================

\section{FSI vs.\ ISS response in $^{10}$He}
\label{sec:fsi-iss}

%===============================================================================

\begin{figure}
\begin{center}
\includegraphics[width=0.43\textwidth]{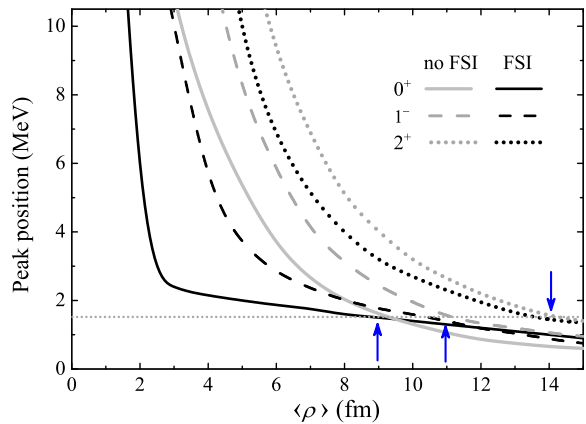}
\end{center}
\caption{Peak positions in the spectrum of $^{10}$He for different $J^{\pi}$ as
function of the source root mean square hyperradius $\langle \rho \rangle$.
Cases of FSI is shown by black lines and without FSI is shown by gray lines.
Arrows mark $\langle \rho \rangle$ typical for sources with different $J^\pi$.
Horizontal dotted line at 1.5 MeV is shown to guide the eye.}
\label{fig:peaks}
\end{figure}

After we realized the importance of the initial state contribution to the
observable spectrum of $^{10}$He populated from $^{11}$Li it is natural to
inquire: ``Where is the borderline between a domination of final state
interaction and an initial state contribution to the observables?'' To answer
this question we have performed systematic model calculations with sources of
different radii. The results of these calculations are summarized in
Fig.~\ref{fig:peaks} providing the peak positions as a function of the source
rms radius.

Let us have a look on the $^{10}$He ground state behavior. For very small source
radii the ground state is not well populated (the first peak in the cross
section is found at high energy). However for $\langle \rho \rangle \sim 3-5$
fm the first peak position is relatively stable at about $2.5-2.0$ MeV. This
energy well corresponds to the $0^+$ resonance position defined in
\cite{Grigorenko:2008} via eigenphases for $3\rightarrow 3$ scattering. With
the source radius increase 
the $0^+$ peak position continue to slide down. At about $\langle \rho \rangle
\sim 9$ fm (for $^{11}$Li $\langle\rho\rangle\sim 8.36$ fm) the FSI curve 
crosses with no FSI curve This probably should be
interpreted in such a way that for such radial characteristics the low-energy
response imposed by ISS can not be any more distinguished from the FSI effect in
principle.

%-------------------------------------------------------------------------------
%-------------------------------------------------------------------------------

%===============================================================================

\section{``Pileup'' of different states in $^{10}$He}
%
%===============================================================================

The Fig.~\ref{fig:peaks} allows to understand qualitatively the picture of the
``pileup'' of states with different $J^{\pi}$, which is shown in
Figs.~\ref{fig:no-fsi-all} (b) and~\ref{fig:spec-j-pi} (b) with and without 
FSI\@. The systematics for different states are very different in Figure 
\ref{fig:peaks}.
For the sources with very large radii both the FSI and no FSI curves crossed the
line with $E_T=1.5$ MeV decay energy at about 9, 11, and 14 fm for $0^+$, $1^-$,
and $2^+$ states respectively (indicated by arrows). However, if we turn to
Fig.~\ref{fig:mean-radii} we find that the
realistic sources for $^{10}$He continuum population have very similar typical
$\langle \rho \rangle$ values corresponding to the expected $q$ values. They are
about 7.8 fm at 140 MeV/c, 10.5 fm at 175 MeV/c, and 13.5 fm at 210 MeV/c for
$0^+$, $1^-$, and $2^+$ states respectively.

Thus the ``pileup'' of states with different $J^{\pi}$
[Figs.~\ref{fig:no-fsi-all}b and~\ref{fig:spec-j-pi}b ] obtained in our model at
energy about 1.5 MeV can be obtained in a very simplistic model
for the source of pure ``geometrical'' effect.
The fact that the peaks of the responses with different $J^{\pi}$  are found at
about the same energy is connected to the systematic increase of the deduced
source radii with the increase of $J$. This is in turn the effect of the
removed recoil particle providing larger radial characteristics to terms of the
source function with higher angular momentum.

%===============================================================================

\section{Limits on the $^{9}$He g.s.}
\label{sec:limits}

%===============================================================================

Strong sensitivity of the observable $^{10}$He spectrum to interactions in the
$^{8}$He-$n$ channel was demonstrated in paper \cite{Grigorenko:2008}.
Exploratory studies of Ref.\ \cite{Grigorenko:2008} were motivated by a
variation in $^{9}$He properties obtained in different experimental studies.
Essential questions here were connected with intensity of the $s$-wave
interaction, which was assumed to be a ``strong'' virtual state in some works
(e.g. Ref.\ \cite{Chen:2001}), and properties of the lowest resonant state
(presumably $p_{1/2}$), which in some works is assumed to have a very small,
evidently not single-particle width (e.g. Ref.\ \cite{Bohlen:1999}). Since that
time the new experimental results on $^{9}$He spectrum continue to be
controversial. On the one hand,  it was found that there is no ``strong''
virtual state in $^{9}$He \cite{Johansson:2010,Falou:2011} providing the
scattering lengths as $a_s \approx -3$ fm and $a_s \geq -3$ fm correspondingly.
On the other hand a low-lying structure, corresponding to $a_s\approx -12\pm 3$
fm was observed
\cite{Kalanee:2013}.

It should be understood that conclusions of \emph{all} the experimental papers
were based on the interpretations which were at the limit of the statistical
reliability or/and the resolution of the corresponding data sets. Because we
think that the reliability of the $^{10}$He spectra obtained in the proton
knockout from $^{11}$Li is confirmed and our understanding of the reaction
mechanism is generally reliable,  we decided to reverse the problem and to check
what are the limitations imposed by the $^{10}$He spectrum on $^{9}$He spectrum
via the theoretical model calculations.

%-------------------------------------------------------------------------------
\begin{figure}
\begin{center}
\includegraphics[width=0.43\textwidth]{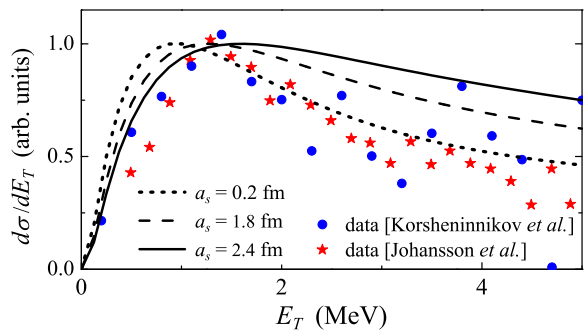}
\end{center}
\caption{Calculation with only $s$-wave interaction included in the $^{8}$He-$n$
channel. The interaction is characterized by the scattering length $a_s$ as
indicated in the legend.}
\label{fig:only-s}
\end{figure}
%-------------------------------------------------------------------------------

For example, to impose limits on the $s$-wave interactions in $^{8}$He-$n$
channel we make calculations with only $s$-wave interactions included in the
three-body Hamiltonian, see Fig.~\ref{fig:only-s}.

The limiting value of the scattering length we define, at the moment, when the
ground state peak goes below about 1 MeV, which make it clearly inconsistent
with the data.

\begin{figure}
\begin{center}
\includegraphics[width=0.43\textwidth]{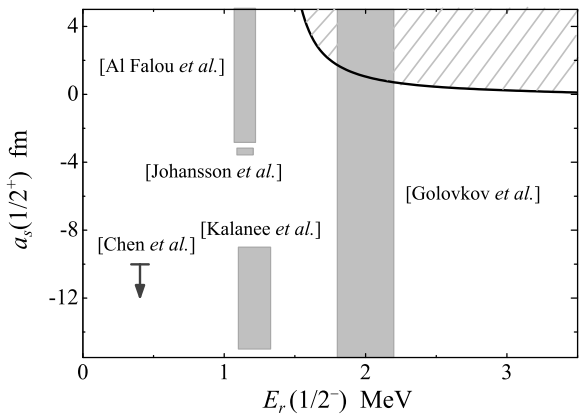}
\end{center}
\caption{Limits on the $s$-wave $1/2^+$ and $p$-wave $1/2^-$ interactions in
$^{9}$He ($^{8}$He+$n$ channel). $a_s$ is the $s$-wave scattering length and
$E_r$ is the $1/2^-$ resonance energy. Hatched area shows the region admissible
according to our calculations. Gray rectangles show experimentally obtained
limits for $^{8}$He+$n$ interactions. Experiment \cite{Chen:2001} gives only
the limit for scattering length. So it is displayed as the arrow.}
\label{fig:s-vs-p}
\end{figure}

If we make such a procedure systematically (requesting the peak position to be
not lower then 1 MeV) also for the $p$-wave interactions, then the constrains
on the plane of $s$-wave vs.\ $p$-wave interaction parameters can be derived,
see Fig.~\ref{fig:s-vs-p}. We can see that practically none of the available
experimental data are consistent with these constrains. The constrains given in
Fig.~\ref{fig:s-vs-p} are, of course, relevant by the nature of our model only
to the single-particle states. However, it is evident that anyhow fully
conclusive experimental studies of $^{9}$He with convincing statistics,
sufficient energy resolution, and clear spin parity identification
of the states are required.

%-------------------------------------------------------------------------------
%-------------------------------------------------------------------------------

%===============================================================================

\section{Three-body correlations}
%
%===============================================================================

The importance of the initial state and recoil effects was clear to the authors
of Ref.\ \cite{Johansson:2010b}, where $^{10}$He was populated in the proton
knockout from $^{11}$Li. They even provide some qualitative estimates of the
expected effects. They suggested a possible ``workaround'' for this problem.
It was shown in \cite{Johansson:2010b} that the three-body correlation patterns
observed for the low-energy peak in $^{10}$He spectrum are different from the
correlations inherent from the momentum correlations of $^{11}$Li WF\@. This
fact was interpreted in such a way, that this peak is not a remnant of initial
$^{11}$Li state, but a true dynamical formation and thus a $^{10}$He ground
state.

%-------------------------------------------------------------------------------
\begin{figure}
\begin{center}
\includegraphics[width=0.25\textwidth]{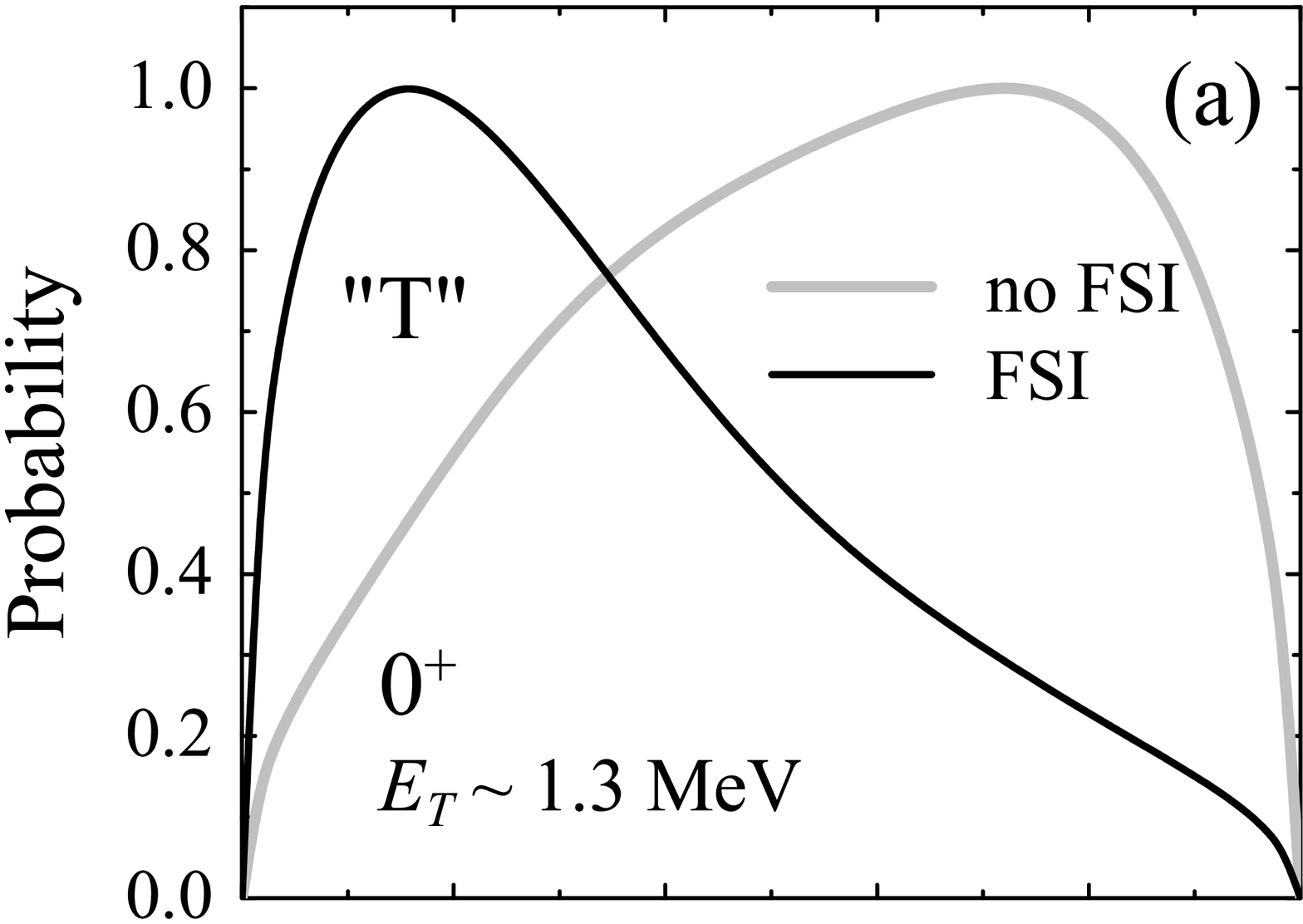}
\includegraphics[width=0.22\textwidth]{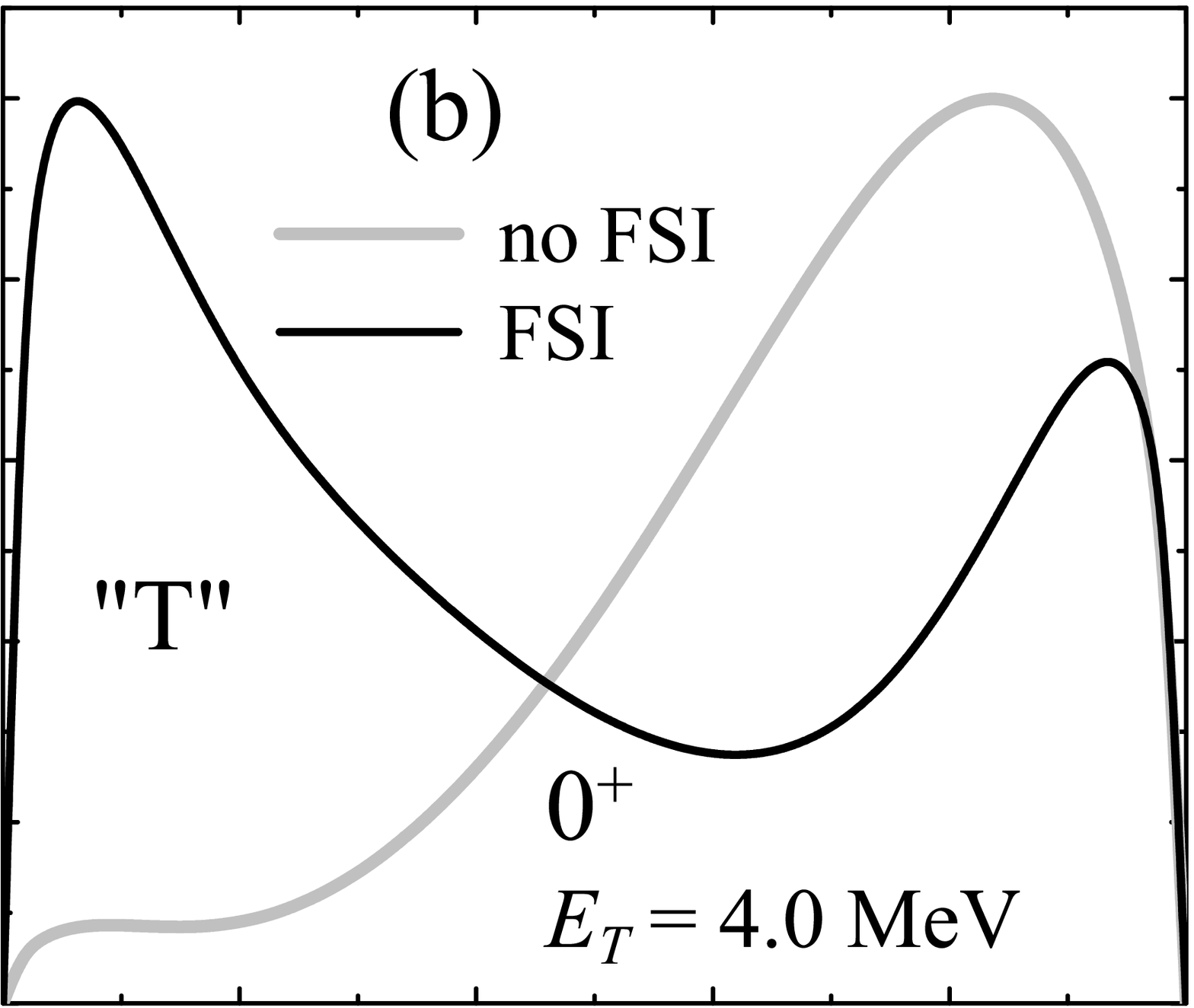}\\
\includegraphics[width=0.25\textwidth]{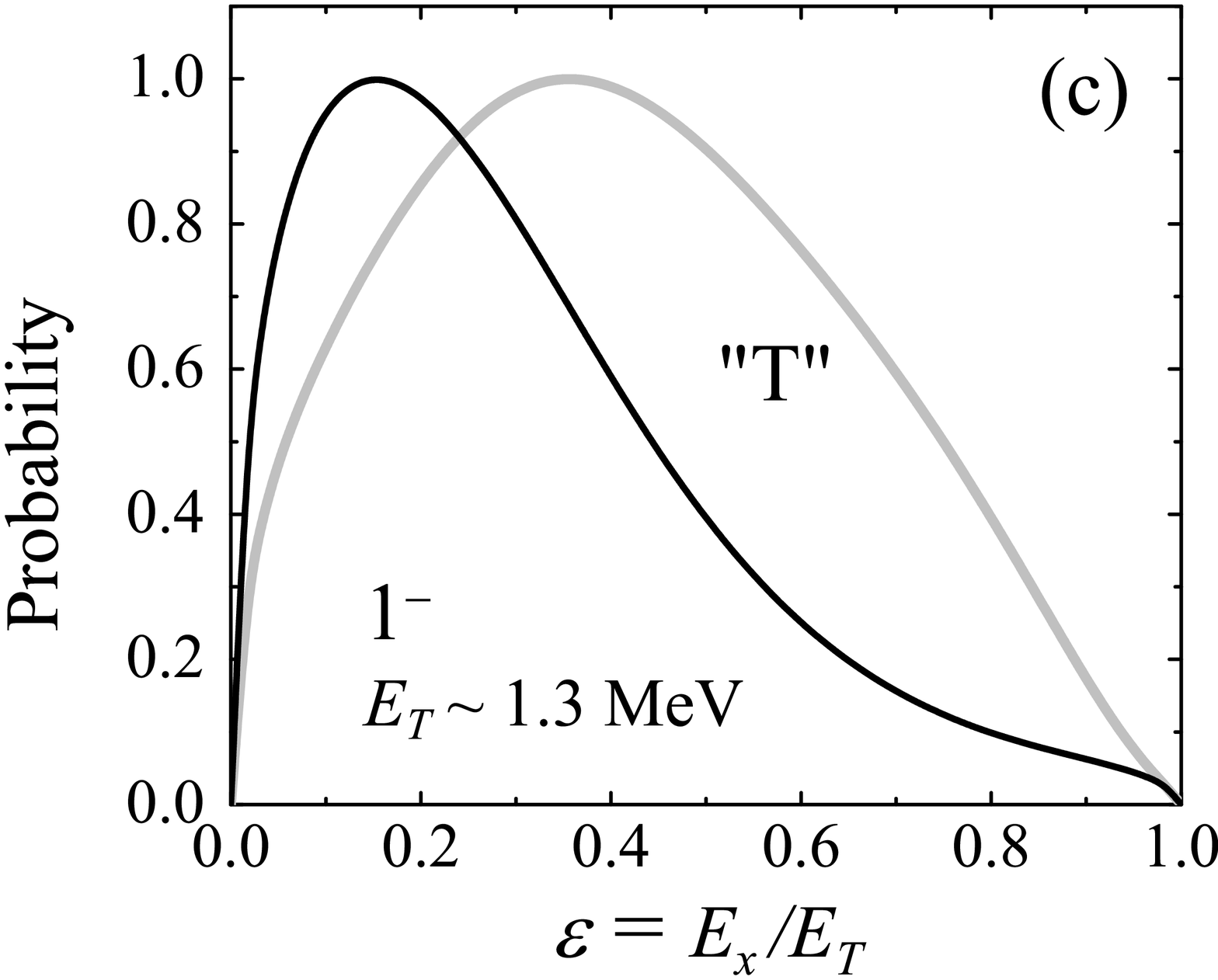}
\includegraphics[width=0.22\textwidth]{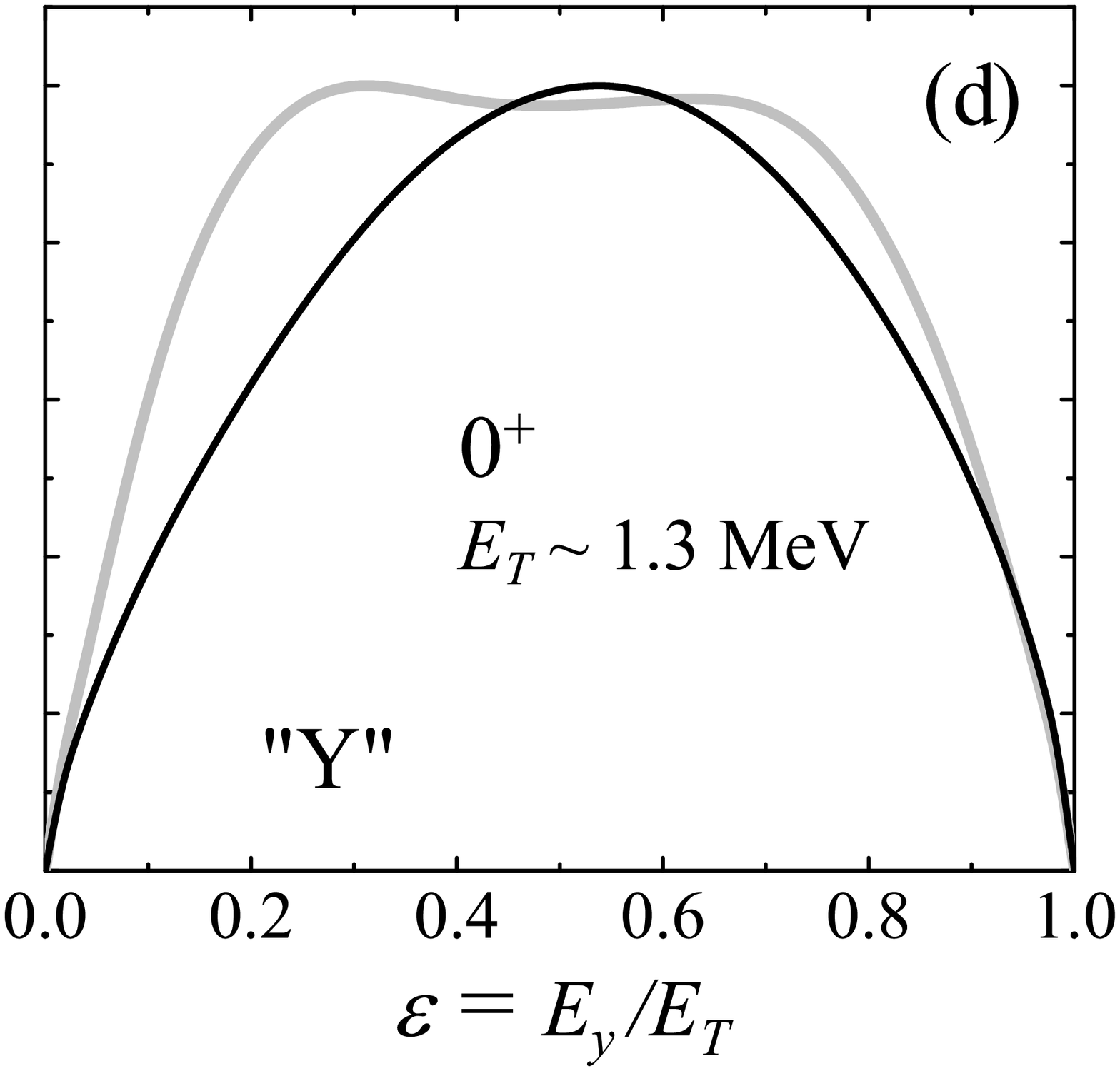}
\end{center}
\caption{Different energy correlations for $^{10}$He populated in proton
knockout from $^{11}$Li. Calculations with and without FSI are shown by black
and gray curves. (a) $0^+$ state, Jacobi ``T'' system, the energy corresponds to 
the
peak value of each distribution. (b)  $0^+$ state, Jacobi ``T'' system, $E_T=4$
MeV. (c) $1^-$ state, Jacobi ``T'' system, energy as in panel (a). (d)
$0^+$ state, Jacobi ``Y'' system, energy as in panels (a) and (c).}
\label{fig:correl}
\end{figure}
%-------------------------------------------------------------------------------

Let us have a look what we get here with much more precise treatment of reaction
mechanism and correct accounting for the $^{10}$He FSIs. Energy distributions
between nucleons are typically most sensitive to decay dynamics.
Fig.~\ref{fig:correl} (a) show this distribution for the peak energy of the
$0^+$ spectrum. The distributions calculated \emph{with} and \emph{without} FSI
demonstrate totally different pictures. The ``no FSI'' distribution is quite
close to the momentum distribution in $^{11}$Li (also shown in
\cite{Johansson:2010b}). However, the full calculation shows a kind of
``opposite'' correlation, evidently dominated by attractive neutron-neutron
contribution. Fig.~\ref{fig:correl} (d) shows also the correlations in the
core-neutron channel. The difference between no FSI and FSI cases is not so
spectacular here, but still quite sizable.

Because the energy spectra calculated \emph{with} and \emph{without} FSI are
very close to each other we arrive to important conclusion about decay dynamics
for this case. In conditions of the powerful ISS effect connected with
$^{11}$Li, the $^{10}$He FSI is not sufficiently strong to modify total
excitation spectrum. However, it is sufficiently strong to rearrange the
correlation patterns. Thus the spectrum shape is based mainly on ISS, while the
three-body correlations are FSI dominated.

The energy range around 4 MeV was assigned to $2^+$ state in Ref.\
\cite{Johansson:2010b}. Important reason for that was the observed change in the
correlation pattern compared to those in the $1-3$ MeV energy range. In our 
calculations we found the $2^+$ contribution small
and focused at small energies. In Fig.~\ref{fig:correl} (b) we demonstrate the
distributions connected with the same $0^+$ excitation but at $E_T=4$ MeV. 
Again, as in Fig.~\ref{fig:correl}
(a) there is a strong difference between no FSI and FSI result. There is also a
strong difference with the FSI distributions of Fig.~\ref{fig:correl} (a) and
(b) calculated for the same $J^{\pi}$. The variation of the energy distributions
Fig.~\ref{fig:correl} (a,b) with energy is therefore not a good indicator for
the change in the $J^{\pi}$. There are an examples, where the problem of the
using the three-body correlations to spin-parity identification was successfully
solved. In Refs.\ \cite{Golovkov:2005,Sidorchuk:2012} this was done for $^{5}$H
and $^{10}$He systems. However, this requires treatment of more complicated
(5-dimensional in general case) correlations connected with orientation of the
three-body system as a whole.

The calculated full distribution in Fig.~\ref{fig:correl} (a) is in a
reasonable agreement with the spectrum from \cite{Johansson:2010b}.
Unfortunately the direct quantitative comparison is not possible here because of
complicated corrections connected with experimental setup. Note, that in
conditions of ``$J^{\pi}$ pileup'' discovered in this work the interference with
the sizable $1^-$ contribution may become important. It has been shown in
\cite{Grigorenko:2012} how the three-body correlation patterns become sensitive
to the interference conditions and experimental bias. The correlations for the
$1^-$ state by itself are given in Fig.~\ref{fig:correl} (c). The demonstrated
energy correlations in $n$-$n$ channel is similar to the energy correlations
for the $0^+$ state. They demonstrate the similar type of the no FSI vs.\ FSI
behavior as the $0^+$ state.

The calculations of this Section demonstrates that conclusions
about $^{10}$He spectrum obtained in Ref.\ \cite{Johansson:2010b} on the basis
of correlations using only speculative argumentation not supported by 
theoretical studies are not valid.

%===============================================================================

\section{Origin of the $1^-$ excitation}
%
%===============================================================================

The evidence for the low-lying $1^-$ state in $^{10}$He was obtained in paper
\cite{Sidorchuk:2012} in the $^8$He$(t,p)$ reaction. It is based on the
observation of asymmetry of the angular distributions of $^{8}$He fragment in
the $^{10}$He c.m.\ frame for $^{10}$He excitation energies $E_T=4-6$ MeV. It
was demonstrated in \cite{Sidorchuk:2012} that such a $1^-$ state position is
consistent with the trend along $N=8$ isotone defined by the shell structure
breakdown in $^{12}$Be. The results of Ref.\ \cite{Sidorchuk:2012} concerning
$1^-$ state were questioned in the paper \cite{Chulkov:2013} basing on
argumentation mainly connected the data treatment and analysis procedures of
Ref.~\cite{Sidorchuk:2012}.

We can comment on the controversy between ideas of \cite{Sidorchuk:2012} and 
\cite{Chulkov:2013} about excited states of $^{10}$He. All the structures 
predicted here for the spectrum of $^{10}$He are quite broad. So, we
immediately can expect strong sensitivity to population mechanism. From
Fig.~\ref{fig:peaks} we can learn that for $1^-$ and $2^+$ states the
sensitivity just to one aspect of reaction mechanism (modeled by the radial
size of the source function) is so strong that it totally overshadows the
predicted effect of the FSI\@. Such structures can not be interpreted as
resonant states (which major properties are independent on the population
mechanism) and should be attributed to so-called ``soft'' excitations (which
appearance is impossible without specific population mechanism).

\begin{figure}
\begin{center}
\includegraphics[width=0.38\textwidth]{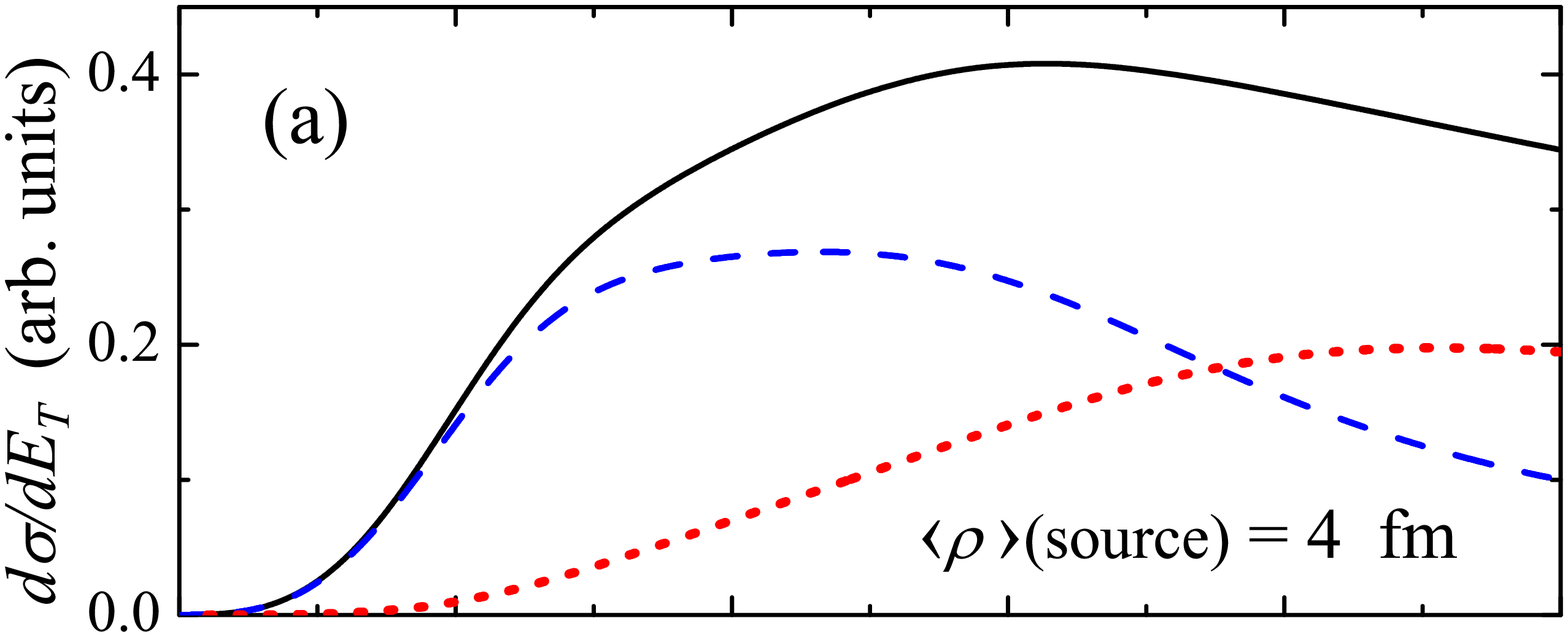} \\
\includegraphics[width=0.38\textwidth]{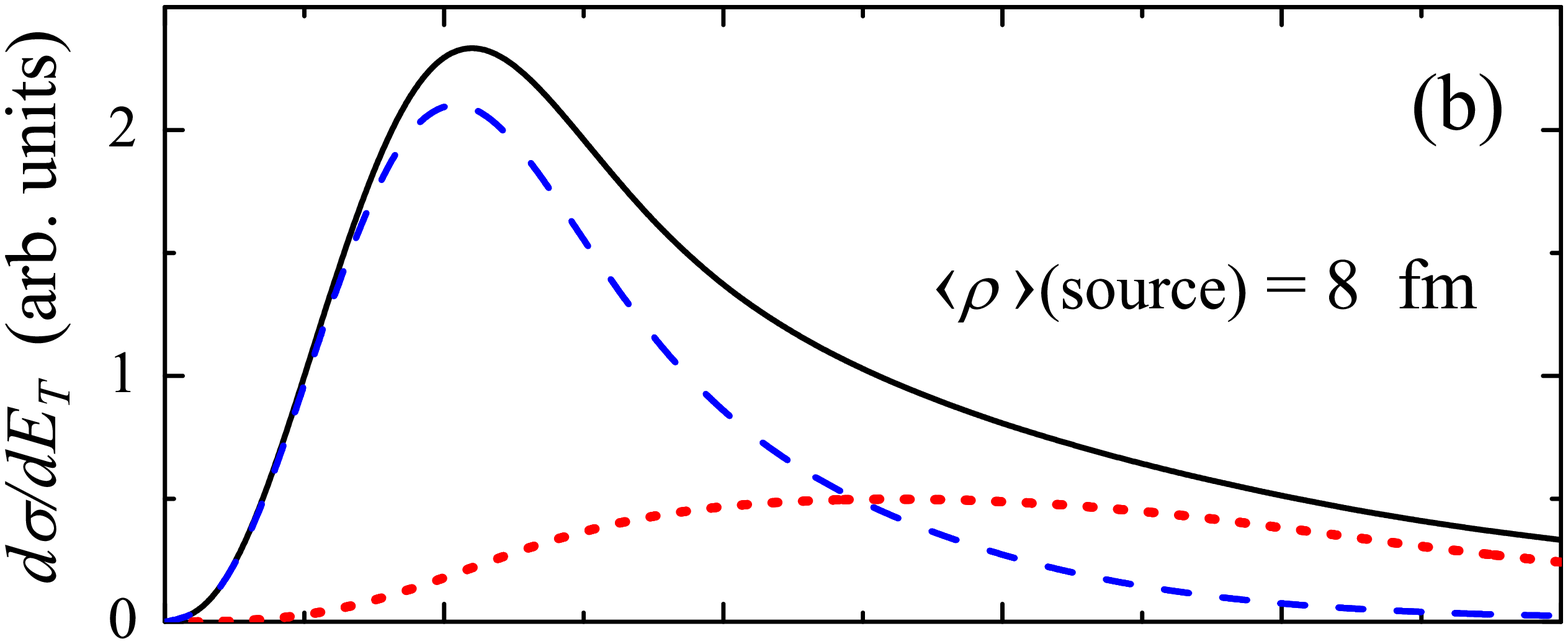} \\
\includegraphics[width=0.38\textwidth]{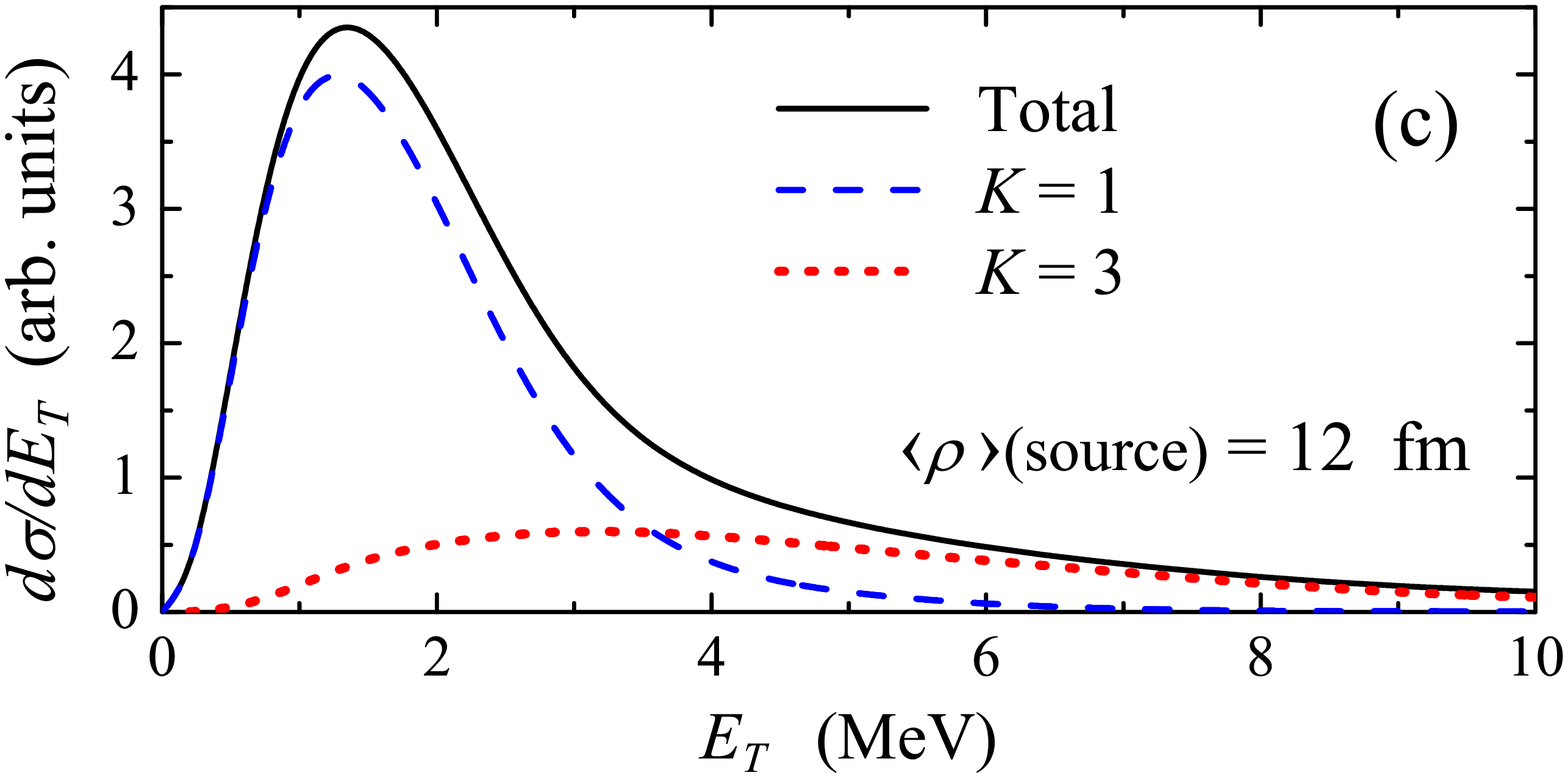}
\end{center}
\caption{Spectra of $1^-$ excitations obtained for different radii of source 
functions $\langle r \rangle = 4$, 8, and 12 fm are shown in panels (a), (b), 
and (c) correspondingly. The contributions of the dominant $K=1$ and $K=3$ 
configurations are given by dashed and dotted curves.}
\label{fig:e1}
\end{figure}

To gain a deeper understanding of the situation we provide in Fig.~\ref{fig:e1} 
selected cases of the spectra used in Figure \ref{fig:peaks}. The partial wave 
decomposition shows that the $1^-$  spectra are always dominated by two 
components of hyperspherical decomposition: $K=1$ and $K=3$. It can be shown 
that in the shell-model-like picture $K=1$ component can be identified with 
$[sp]$ configuration, while $K=3$ is closely related to $[dp]$. The low-energy 
part of spectrum is always connected with $[sp]$ configuration. In such a 
configuration one particle is in resonance ($p$-wave) and one in non-resonant 
($s$-wave), so this is real ``soft'' excitation totally depending on initial 
configuration. In contrast, for the $[dp]$ configuration resonant behavior
could be expected in principle (based on the $p$- and $d$-wave states of
$^{9}$He), but do not really show up in calculations. The profiles for
population of the $[sp]$ and $[dp]$ configurations are very different for
different calculations. While for experimental conditions of Ref.\
\cite{Johansson:2010} practically only $[sp]$ configuration is expected to
contribute the spectrum, for conditions of Ref.\ \cite{Sidorchuk:2012} the
large mixing of configurations is predicted.

Our message here is that not only it is very probable to expect strong 
population of $1^-$ continuum at $E_T=4-6$ MeV in experiment 
\cite{Sidorchuk:2012}, it is also likely that the data \cite{Johansson:2010} of 
the Authors of Ref.\ \cite{Chulkov:2013} itself contains important $1^-$ 
contributions. We should remind here again that our conclusions are valid only
for the $^{10}$He states ``built'' on the single-particle structures of
$^{9}$He subsystem.

%-------------------------------------------------------------------------------
%-------------------------------------------------------------------------------

%===============================================================================

\section{$^{10}$He production in reactions with $^{14}$Be}
%
%===============================================================================

A novel way to populate $^{10}$He was used in the recent work 
\cite{Kohley:2012}. The $^{8}$He+$n$+$n$  invariant mass was reconstructed for 
$^{14}$Be beam reaction on the light target. The $^{10}$He  spectrum well 
coincide (within errorbars) with the spectra 
\cite{Korsheninnikov:1994,Johansson:2010}. However, somewhat different 
properties of $^{10}$He g.s.\ are inferred with $E_T = 1.6(0.25)$, 
$\Gamma=1.8(4)$ MeV. This experiment, presented by the Authors as the 
``two-proton removal'' has two most probable interpretations. One
interpretation is the two-proton knockout populating continuum of $^{12}$He
system. Broad continuum states of the unknown $^{12}$He are then decaying
``sequentially'' populating the $^{10}$He ground state. Alternative
opportunity, which we also find very probable, is the $\alpha$ knockout.

\begin{figure}
\begin{center}
\includegraphics[width=0.4\textwidth]{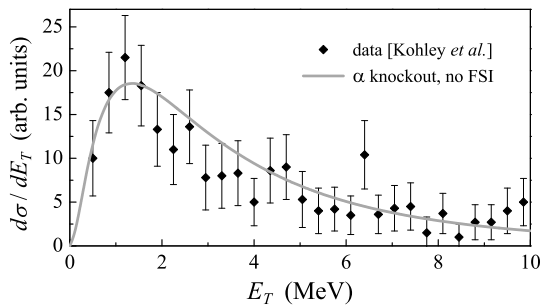}
\end{center}
\caption{``No FSI'' estimate for the $^{10}$He spectrum populated in $\alpha$
knockout from $^{14}$Be. The $q=100$ MeV/c value is selected. }
\label{fig:kohley}
\end{figure}

Let's discuss the second opportunity. The $\alpha$ knockout from $^{14}$Be can 
be treated in the same model as we applied for the knockout from $^{11}$Li. We 
then $^{12}$Be is considered as a core for $^{14}$Be. This is not an extremely 
well defined core, however we can find the $^{14}$Be three-body 
$^{12}$Be+$n$+$n$ WF in the literature \cite{Forssen:2002}. The $^{12}$Be core 
we consider as a $^{8}$He+$\alpha$ system  in a potential model providing a 
reasonable size and correct separation energy for alpha particle. The estimates 
without FSI shown in Fig.~\ref{fig:kohley}, demonstrate a qualitative agreement 
with the data and indicate that the large ISS effect analogous to that in 
$^{11}$Li case can be taking place in the case of the halo structure of 
$^{14}$Be. It seems that the more sophisticated calculations are not necessary 
at the moment until the experimental situation becomes more definite.

%-------------------------------------------------------------------------------
%-------------------------------------------------------------------------------

%===============================================================================

\section{Conclusions}

%===============================================================================

We would like to emphasize the following main results of this work

\noindent (i) The calculations of this work indicate that a ``ground state''
peak observed in the spectrum of $^{10}$He populated in reactions with $^{11}$Li
is likely to be a superposition of $1^-$, $0^+$, and $2^+$ states.  The
calculated effect could be seen as one of the most powerful indications of the
abnormal halo size of $^{11}$Li nucleus.

\noindent (ii) ``Pileup'' of $1^-$, $0^+$, and $2^+$ states obtained in the
calculations of this work is a consequence of the systematic increase of the
radial sizes of sources with $J$. This is, in a row, consequence of the core
recoil effect in the case of expressed halo structure.

\noindent (iii) The energy transfer to $^{10}$He is small in all considered
model assumption if $^{11}$Li WF as initial state is involved. In contrast, the 
angular momentum transfer to $^{10}$He considerably depends on the properties
of $^{11}$Li WF.

\noindent (iv) The actual lowest excitation in the $^{10}$He spectrum is
predicted to be ``soft'' $1^-$ peak, created by an extreme spatial extent of the
initial state configuration ($^{11}$Li halo nucleus). This is an extremely
exotic situation having no match so far in the other systems and reactions.

\noindent (v) The strong initial state effect could be a generic problem for
reactions populating broad states of $2n$ emitting systems beyond the driplines.
We can mention here recently measured $^{13}$Li populated in the proton knockout
off $^{14}$Be \cite{Johansson:2010} and $^{16}$Be populated in the proton
knockout off $^{17}$B \cite{Spyrou:2012}.

\noindent (vi) Our calculation result reconcile the experimental results 
obtained
in the knockout reactions on halo systems
\cite{Korsheninnikov:1994,Johansson:2010,Kohley:2012} with the data from the
($t$,$p$) transfer reaction \cite{Golovkov:2009,Sidorchuk:2012}. In our
interpretation there is no problem with any of these data by itself, but the
interpretations should be very different. This new interpretation prescribe
``new'' ground state position for $^{10}$He $\sim 2.1$ MeV \cite{Sidorchuk:2012}
compared to value $\sim 1.3$ MeV accepted since discovery of $^{10}$He
\cite{Korsheninnikov:1994}.

\noindent (vii) The data on $^{10}$He spectrum can be used to impose constrains
on the spectrum of $^{9}$He. These constrains are valid under the assumption of
a single-particle nature of the low-lying states of $^{9}$He. Within this
assumption we see that practically all the data on the $^{9}$He $1/2^+$
and $1/2^-$ states are incompatible with the $^{10}$He data discussed in this
work.

\noindent (viii) Three-body correlation studies further clarifies the role of 
the
final state interaction in the situations discussed here. We have found in this
work that the $^{10}$He FSI is insufficiently strong to modify the excitation
spectrum, governed thus by the initial state structure. However, it is strong
enough to completely rebuild the correlations patterns of the fragments.

Situation with the studies of the $^{10}$He g.s.\ which we address in this work
underline the importance of the theoretical studies in interpretation of complex
and unusual phenomena in the exotic dripline systems. Our results also call for
deeper and finally more conclusive experimental studies of $^{10}$He system.
Such studies should make $^{10}$He benchmark system and basis for understanding
of the two-neutron emitters beyond the dripline.

%===============================================================================
%
\textit{Acknowledgments.}
%
%===============================================================================
%
--- The work was carried out with the financial support of SAEC ``Rosatom'' and 
Helmholtz Association.
%P.G.S.\ and I.A.E.\ are supported by the Helmholtz Association under grant
%agreement IK-RU-002. 
L.V.G.\ is supported by RFBR 11-02-00657-a grant and
Russian Ministry of Industry and Science grant NSh-932.2014.2.

%###############################################################################

%###############################################################################

\end{document}